\documentclass[prd,twocolumn,superscriptaddress,nofootinbib,10pt]{revtex4-1}
\pdfoutput=1
\usepackage[utf8]{inputenc}
\usepackage{xcolor,cancel}
\usepackage{subfigure}
\usepackage{braket}
\usepackage{epsfig} 
\usepackage{amsmath} 
\usepackage{amsfonts}  
\usepackage{float}    
\usepackage{amssymb}   
\usepackage{slashed} 
\usepackage{color}       
\usepackage{pbox}
\usepackage{array,multirow,makecell}
\usepackage[colorlinks,citecolor=blue, linkcolor = blue, urlcolor = blue]{hyperref}
\usepackage[normalem]{ulem}
\usepackage{tabularx}
\usepackage{titlesec} 
\usepackage{scrextend}
\usepackage{natbib} 
\usepackage{array,multirow,makecell}
\usepackage{lipsum}
\usepackage{mathtools}

\newcommand{\lsim}[1]{\lesssim}
\newcommand{\bew}{\begin{widetext}}
\newcommand{\eew}{\end{widetext}}
\begin{document}
\begin{flushright}
MI-HET-850
\end{flushright}
\title{Probing Light Particles With Optically Trapped Sensors Through Nucleon Scattering}

\author{Bhaskar Dutta}
\email{dutta@tamu.edu}
\affiliation{Texas A$\&$M University, College Station, Texas 77843, USA}

\author{Dilip Kumar Ghosh}
\email{tpdkg@iacs.res.in}
\affiliation{School of Physical Sciences, Indian Association for the Cultivation of Science,\\ 2A $\&$ 2B Raja S.C. Mullick Road, Kolkata 700032, India}

\author{Sk Jeesun}
\email{skjeesun48@gmail.com}
\affiliation{School of Physical Sciences, Indian Association for the Cultivation of Science,\\ 2A $\&$ 2B Raja S.C. Mullick Road, Kolkata 700032, India}

\begin{abstract}
Optically levitated nanospheres are highly sensitive to the motion of their center of mass even under small momentum transfer. We propose detecting exotic particles via nucleon scattering in such spheres in the context of an ongoing experiment. The 200 nm-diameter spheres within the present experimental realization, featuring a configuration of the array $4\times 4$ and its upgrade, can achieve sensitivity to nuclear couplings of ALPs exclusively and pseudoscalar dark matter in the $\sim 10$ keV mass range, targeting previously unconstrained regions of parameter space. 
In contrast, a smaller sphere with a diameter of 15 nm benefits from overall coherence enhancement, enabling the detection of pseudoscalar and vector dark matter down to $\mathcal{O}(100)$ eV even with a single sphere.
This smaller setup also offers the potential for the direct detection of Earth-bound dark matter strongly coupled with visible matter, even with its minimal velocity and tiny fractional abundance.
\end{abstract}
\maketitle
The elucidation of dark matter (DM) evidence 
constitutes a significant theoretical and experimental challenge within the field of high-energy physics \cite{Cirelli:2024ssz}.
The null results for the larger than GeV scale weakly interacting DM from direct search experiments have motivated us to look for it in the light mass 
range. Apart from DM there also exist several well-motivated beyond the Standard Model (BSM) scenarios which include {\it axion} and {\it axion-like particles} (ALP) \cite{Peccei:1977hh, Weinberg:1977ma, Wilczek:1977pj,Kim:2008hd,Caputo:2024oqc}.
The detection of such light particles is difficult in conventional direct detection (DD) experiments because of the threshold requirements in recoils.  
There are theories predicting strongly interacting dark matter (DM) particles that might evade detection in underground DD experiments. These particles lose energy rapidly through multiple scatterings, preventing them from reaching detectors. Over time, such interactions can lead to a significant accumulation of DM density near Earth's surface~\cite{Zaharijas:2004jv, Neufeld:2018slx, Bramante:2022pmn, McKeen:2023ztq}. This phenomenon, known as ``Earth-bound DM", presents unique challenges for conventional detection methods.

Recently, much attention has been paid to develop new experimental facilities with lower threshold energy to detect such light exotic particles by looking at electron ionization \cite{Essig:2015cda} or excitation \cite{Hochberg:2017wce}, placing new limits on interactions with photons and electrons \cite{XENON:2024znc,DAMIC-M:2023gxo,SENSEI:2024yyt}. 
However, the nucleon couplings of these light particles remain difficult to investigate~\cite{XENON:2023cxc,LZ:2022lsv,PandaX:2023xgl,Dutta:2024kuj}.
Optical monitoring of the recoil of levitated nanospheres in new tabletop experiments is increasingly being used to detect light BSM particles~\cite{Riedel:2012ur, Monteiro:2020wcb,Moore:2020awi, Carney:2022pku, Afek:2021vjy}. 

As proposed in~\cite{Afek:2021vjy}, the ongoing experiment utilizes
a single nanosphere with a diameter of (100-200) nm. However, current advances have already achieved a $4\times 4$ array of such nanospheres \cite{Yan:23} and its enhancement to a array size of $\mathcal{O}(10)\times \mathcal{O}(10)$ \footnote{Private communication with David Moore.}.
These optically trapped sensors are highly sensitive to the center-of-mass motion of the nanospheres, even at noise levels that approach the standard quantum limit (SQL)~\cite{Caves:1981hw}.
It has been proposed in ref.~\cite{Afek:2021vjy} that with SiO$_2$ material, the setup with a single nanosphere is most sensitive to (three) momentum recoil of $q\sim \mathcal{O}(10)~{\rm keV}- {\rm few}~100$ keV for a sphere of 200 nm diameter (femtogram) and $q\sim 85$ eV $-\mathcal{O}(100)~$ eV for a 15nm diameter sphere (attogram).
The lower sensitivity limit is determined by the threshold, specifically the SQL, while the
form factors of the nanospheres control the upper limit, as discussed in the following. 
In comparison, the existing DD experiments are sensitive to momentum transfer $q\sim \sqrt{2 m_N E_R}\gtrsim $ few MeV for nuclear recoil~\cite{XENON:2020rca}.
Recent experimental advances in other systems have demonstrated the ability to surpass SQL 
\cite{Gross:2010rsg,Hosten:2016yho,Rossi:2018dav,McCuller:2020yhw}  and similar techniques have been proposed for levitated sphere setups. These innovations could improve sensitivity, enabling the detection of even smaller recoil signals \cite{Afek:2021vjy,Kilian:2024fsg}.
 Making large arrays 
 ($\sim1000\times 1000$ 2D arrays) in the future of such optically trapped sensors may enable further improvements in the search for light DM as well as other BSM particles \cite{Afek:2021vjy} and some of the larger arrays (e.g. $6100$ 1D arrays)  have already been achieved \cite{Zhang:2017kde,Moore:2020awi,Gilmore:2021qqo,Manetsch:2024lwl}.
With these arrays and very low momentum sensitivity, levitated nanospheres can be extremely helpful in probing low-energy nuclear recoils.

In this {\it Letter}, we propose using the levitated nanosphere setup to detect nuclear interactions of light BSM particles.   
We aim to probe ALPs utilizing the 14.4 KeV line flux of solar ALPs from $^{57}$Fe de-excitaion \cite{CUORE:2012ymr} and pseudo-scalar and vector DM utilizing galactic halo DM flux. 
We will demonstrate that the levitated nanosphere setup offers a powerful tool to explore a substantial portion of previously inaccessible parameter space. In particular, 
the $4\times 4$ array of large nanospheres, has already begun probing unrevealed regions of parameter space for ALP and DM. Furthermore, we investigate the detection prospects for GeV-scale Earth-bound DM with extremely low kinetic energy ($\sim 0.03$ eV) which may evade conventional DM detectors~\cite{McKeen:2023ztq}.
Although alternative approaches have been proposed to investigate such scenarios~\cite{Neufeld:2018slx,McKeen:2023ztq,Ema:2024oce,Pospelov:2019vuf}, the levitated nanosphere setup could offer a direct and robust detection method for this kind of DM.

{\it Scattering rate.} The differential recoil rate of a particle $\phi$ in such a single nanosphere setup is given by $
\frac{dR}{dq}=  \int dE_\phi \frac{d\Phi_\phi}{dE_\phi}\frac{d\sigma}{dq} ~S(q), $ 
where $R, E_\phi$ and $\Phi_\phi$ are the recoil rate, energy, and incoming flux of $\phi$ respectively. $S(q)$ is the structure factor that contains the details of the target material and is given as \cite{Afek:2021vjy},~
$S(q)= \left[\sum_i N_i Z_i^2 F_H^2(q r_{A_i}) + N_p^2 F_c^2(q)   \right]$.
$N_i$ denotes the number of different nuclei in the target and the number of protons and neutrons (depending on the coupling of $\phi$) in each nucleus is denoted by $Z_i$. 
$r_{A_i}=1.22\times A_i^{1/3}$ fm is the nuclear radius with mass number $A$. 
$F_H$ is the Helm form factor \cite{Lewin:1995rx}.
The first term in $S(q)$ dominates in larger $q$ ($\sim$ 10-100 keV), contributing to coherent scattering from the nuclei in larger spheres (e.g., 200-nm diameter). In contrast, the second term becomes significant at lower $q~(\ll 2\pi/r_{sp}$ $\lesssim 0.1$ keV), where coherent scattering involves the entire sphere. This behavior is a characteristic of smaller spheres (e.g., 15 nm diameter) and is governed by the form factor $F_c(q)= 3 j_1(r_{sp} q)/(q r_{sp})$.
Here, $r_{sp}$ and $N_p$ signify the radius of the sphere and the total number of protons and neutrons (depending on the coupling of $\phi$) in the sphere, respectively. 
The smaller nanospheres can be sensitive to momentum up to $q\sim$ few keV. However, for $0.1$ keV$\lesssim q\lesssim 100$keV,  coherence is lost and the sensitivity scales with the number of nuclei
 $\sim \sum_i N_i Z_i^2$. 
Smaller nanospheres are still useful for probing momentum transfers up to a few keV. Once the threshold, a few keV,  is surpassed, larger nanospheres (e.g., 200 nm diameter) are more advantageous because of their $\sim 10^3$-fold greater number of target nuclei compared to smaller ones (e.g., 15 nm diameter).

In the following, we organize the discussion of light particle detections based on the recoil momentum of the incoming particles and emphasize the parameter spaces relevant to various physics scenarios that benefit from sensitivity to different recoil-momentum ranges.

{\bf I.  Large nanosphere ($ q \sim \mathcal~{O}(10)~{\rm keV} $):}
By the {\it large} sphere, we refer to the optically trapped spheres of $200$ nm ($r_{sp}=100$ nm) diameter. From the discussion in the introduction, it is clear that if the incoming particle has roughly $q\sim \mathcal {O}(10) $ keV momentum and $> 2\pi/r_{sp}$, it can scatter with individual nucleons in the sphere. Thus, for such particles, we need a large number of nucleons with momentum sensitivity as low as $\sim$ few tens of keV, and hence we use the {\it large} sphere for such particles. 
We consider SiO$_2$ nanosphere with $200$nm diameter optically trapped in a harmonic potential with frequency $\omega=2\pi\times 20$ KHz \cite{Afek:2021vjy}. Si (O) has a mass number $28$ (16) and the density of SiO$_2$ is $1.8$ gm/cm$^3$. The SQL for momentum impulses for this sphere is $\sigma_{SQL}=\sqrt{m_{sp} \omega}=1.8\times 10^4$ eV. We now explore different scenarios in the following subsections.

I.A {\it Solar ALP (non-dark matter):}
{\it Axions} are the pseudo-Nambu Goldstone bosons (pNGB) associated with the spontaneously broken global symmetry $U(1)_{PQ}$ (also known as {\it Peccei-Quinn} symmetry) and was proposed to solve the strong CP problem of QCD \cite{Peccei:1977hh, Weinberg:1977ma, Wilczek:1977pj,Kim:2008hd}.
However, pNGBs may also appear in several BSM paradigms that are not related to the strong CP problem and are generally known as ALP.  
Axions and ALPs are being investigated in numerous ongoing and future experiments with the assumption of some effective coupling of ALP with electron ($g_{ae}$), 
photon ($g_{a\gamma}$) or nucleon ($g_{aN}$) \cite{Caputo:2024oqc}. 
Among these couplings, the constraint on $g_{aN}$ is obtained primarily in combination with $g_{ae}$ and $g_{a\gamma}$ \cite{CAST:2017uph,Lucente:2022esm,Capozzi:2023ffu} since it is difficult to produce and detect ALP via a coupling $g_{aN}$. Sun can emit ALPs with energy around both the $\sim$ keV and the MeV scale through nuclear de-excitations \cite{CUORE:2012ymr,Bhusal:2020bvx}, Compton \cite{Derbin:2011gg} or Primakoff processes \cite{vanBibber:1988ge}.
DM detection experiments act as very sensitive probes of $g_{ae}$ or $g_{a\gamma}$ as the electronic excitation is sensitive to $\sim$ keV recoil energy \cite{PandaX:2017ock,XENON:2020rca}.
However, for the ALP-nucleon interaction, current direct detection experiments face the challenge of detecting very low-energy nuclear recoils \cite{PandaX:2017ock,XENON:2020rca}. 
Previously, the coupling $g_{aN}\gtrsim 4\times 10^{-5}$ was constrained using a limited solar ALP flux with energy
$\sim 5.5$ MeV,  as observed in the Sudbury Neutrino Observatory (SNO) experiment \cite{Bhusal:2020bvx}. In this work, our objective is to investigate
$g_{aN}$ independently, relying solely on the 14.4 keV line.

To probe the ALP nucleon coupling we consider the following Lagrangian,
$    \mathcal{L}_{aN}\supset -i  \Bar{N}\gamma^5 (g^0_{aN} \mathbb{I} + \sigma_3 g^3_{aN}) N a, $
where, $N=(n,p)$ neutron proton doublet and $g^0_{aN},~g^3_{aN}$ are the iso-scalar and iso-vector couplings, respectively.
The couplings with proton ($g_{ap}$) and neutron ($g_{an}$) are defined as $g^0_{aN}-g^3_{aN}$ and $g^0_{aN}+g^3_{aN}$ respectively.
Sun can emit ALP with energy $14.4$ KeV from deexcitation of the $^{57}$ Fe nucleus and the flux $d\Phi_a/dE_a$ is discussed in detail
in ref.\cite{CAST:2009jdc, CUORE:2012ymr}.
Upon reaching Earth, these ALPs scatter with the nucleons in the optically trapped nanospheres. 
One such process that causes nuclear recoil is $a (k_1)+N(k_2)\to \gamma (k_3)+N (k_4)$ and the amplitude will be similar to that of pion photoproduction \cite{Guo:2022hud}. 
Since we are dealing with solar ALPs with $\sim$ keV energy (thus $(q\lesssim{\rm MeV})$), the differential cross section $\frac{d\sigma}{dq}$ becomes identical to the one for $a+e\to e+\gamma$ with obvious replacements \cite{Avignone:1988bv,Vergados:2021ejk}.
The differential event rate per trapped sphere is given by,
\begin{equation}
\frac{dR}{dq}=  \left.\frac{d\Phi_a}{dE_a}\right\vert_{E_a=E_0} \left.\frac{d\sigma}{dq}\right\vert_{E_a=E_0}~\Theta(E_a-E_a^{\rm min}) ~S(q) 
\label{eq:ratea}
\end{equation}
where $E_a^{\rm min}$ is the minimum incoming energy kinetically required to generate a recoil momentum $q$. $E_0=14.4$ keV is the incoming ALP energy.
%
\begin{figure}[!tbh]
    \centering
    \includegraphics[scale=0.35]{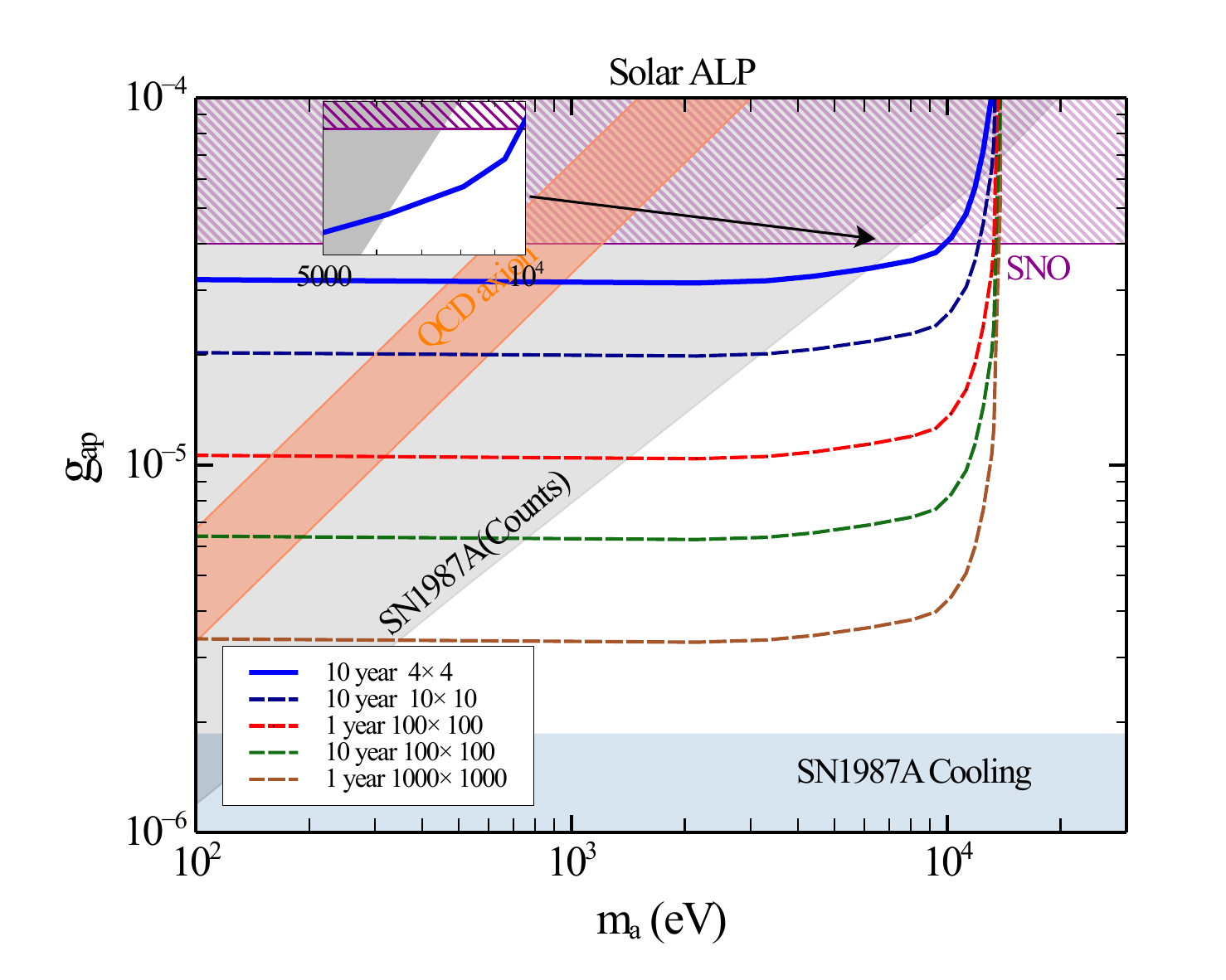}
    \caption{Projected sensitivity at $90\%$ C.L on the $m_a - g_{ap}$ plane for solar ALP using different combination of levitated SiO$_2$ nanosphere array with $200$ nm diameter and momentum threshold $0.5~\sigma_{\rm SQL}$ (discussed in the text). Shaded regions are existing constraints.}
    \label{fig:solar_alp}
\end{figure}
Nuclear coherence scattering occurs because the solar ALP has energy $E_a\sim 14.4 $ keV.
Since the flux energy is slightly less than $\sigma_{SQL}=18$ keV we use $0.5 \sigma_{\rm SQL}$ as the threshold \cite{Afek:2021vjy}.
We assume  negligible  background $(\ll 1) $ for the chosen threshold \cite{Afek:2021vjy} to obtain constraints.
 The sensitivity at $90\%$ C.L. obtained for the detection threshold 
$q_{th} = 0.5 \sigma_{\rm SQL}$ in the $m_a - g_{ap}$ plane is shown in 
Fig.\ref{fig:solar_alp}.
We use different colored lines to indicate the limit obtained for
varied configurations of array size and integration time (10 year $4\times 4$ (blue solid),10 years $10\times 10$ (blue dashed), 1 year $100\times100$ (red dashed), 10 years $100\times 100$ (green dotted), 1 year $1000\times 1000$ (brown dashed dot)).

We find that even the $4\times 4$ array setup of the levitated sphere can probe $g_{ap}\sim 3 \times 10^{-5}$ while the $100\times 100$ array for 1 year (10 year)  
experiments can probe  $g_{ap}\sim 10^{-5}~(7 \times 10^{-6})$  in the low mass range ($m_a\lesssim 10$ keV).
The only experimental constraint relevant in the parameter space is from SNO \cite{Bhusal:2020bvx}.
With the $1000\times1000$ array, it can be sensitive even up to $g_{ap}\sim 3 \times 10^{-5}$.
Astrophysical constraints such as SN1987A counts \cite{Lella:2023bfb} and SN1987A cooling \cite{Carenza:2019pxu} are also relevant in the parameter space shown by gray and light blue shaded regions, respectively.
However, it is interesting to note that the formalism of obtaining supernova bounds is associated with astrophysical uncertainties \cite{Bar:2019ifz}. 
The limit obtained using
nanospheres is complementary to the SN1987A counts bound for $m_a<$ keV and thus would shed light into these previously unexplored regions.
In regions of very low mass ($m_a\ll 10^{-2}$ eV) the limit can be better compared to other experimental searches
of ALP nucleon coupling looking for DM ALPs such as CASPER \cite{Wu:2019exd}, SERF comagnetometer \cite{Bloch:2022kjm}.
There also exist experimental limits and future projections using solar ALPs to provide constraints on the product of $g_{ae}\times g_{aN}$ \cite{Borexino:2012guz,Lucente:2022esm} or $g_{a\gamma}\times g_{aN}$ \cite{CAST:2009jdc,Armengaud:2013rta,Lucente:2022esm}.
However, our projections in Fig.\ref{fig:solar_alp} are more robust as they constrain individual
nucleon coupling.
These limits can also probe the parameter space of the QCD axion shown by the orange band in Fig.\ref{fig:solar_alp}.
As noted previously, solar ALPs can also be probed using a $15$ nm diameter sphere. However, the
$\sim 10^3$-fold reduction in the number of target protons compared to larger spheres weakens the sensitivity limits by approximately a factor of $\sim6$.

I.B {\it Sub-GeV Pseudo-scalar DM:}
\label{subsec:alpdm1}
The sub-GeV scale particles like scalar, pseudo-scalar  or vector are well motivated DM candidates and are widely being studied \cite{CRESST:2017ues,XENON:2020rca}. 
As an example, here we discuss the detection prospects of sub-MeV pseudo-scalar DM ($\chi_s$) in large nanospheres.
$\chi_s$ can be produced through different mechanisms and can saturate the observed density of DM \cite{Abbott:1982af,Dine:1982ah}. 
Since cold DM at present is non-relativistic $(v\sim 10^{-3}c)$, it can be absorbed into target materials through the axio-electric effect ($\chi_s+e\to e+\gamma$) with a mono-energetic signal at $q\sim m_{\chi_s}$. This process allows constraints to be placed on the electron-DM coupling $g_{e\chi_s}$ from Xenon 1T \cite{XENON:2024znc}, SuperCDMS \cite{SuperCDMS:2019jxx},EDELWEISS \cite{EDELWEISS:2018tde}.
However, as mentioned earlier, the current direct detection experiments lose sensitivity
for a low mass ($\lesssim$GeV) dark matter nuclear scattering due to their high detection threshold leading to a lower constraint on the proton-DM coupling $g_{p\chi_{s}}$.
 The levitated sphere can probe $g_{p \chi_s}$ through the nuclear scattering process: $p + \chi_s \to p + \chi_s$ calculated using the
following effective Lagrangian: $ \mathcal{L}_{\chi_{s} p}\supset -i g_{p\chi_{s}}\Bar{p}\gamma^5 p\chi_s$.
This analysis assumes stable dark matter without specifying
its production mechanism. The recoil rate for pseudo-scalar DM absorption with mass $m_{\chi_s} $ will be  \cite{Dror:2019onn,Dror:2019dib},
  $  R=\frac{\rho_{\rm DM}}{m_{\chi_s}} \sigma_{\chi_{s} p}~ S(q=m_\chi) \Theta (m_\chi-q_{th}) $, 
with $\sigma_{\chi_s p}$ being the absorption cross-section per proton \cite{Dror:2019dib} and $S(q)$ being the structure factor mentioned earlier. 
We consider the DM halo density $\rho_{\rm DM}=0.3 ~{\rm GeV}/{\rm cm}^3$.

\begin{figure}[!tbh]
    \centering
    \includegraphics[scale=0.35]{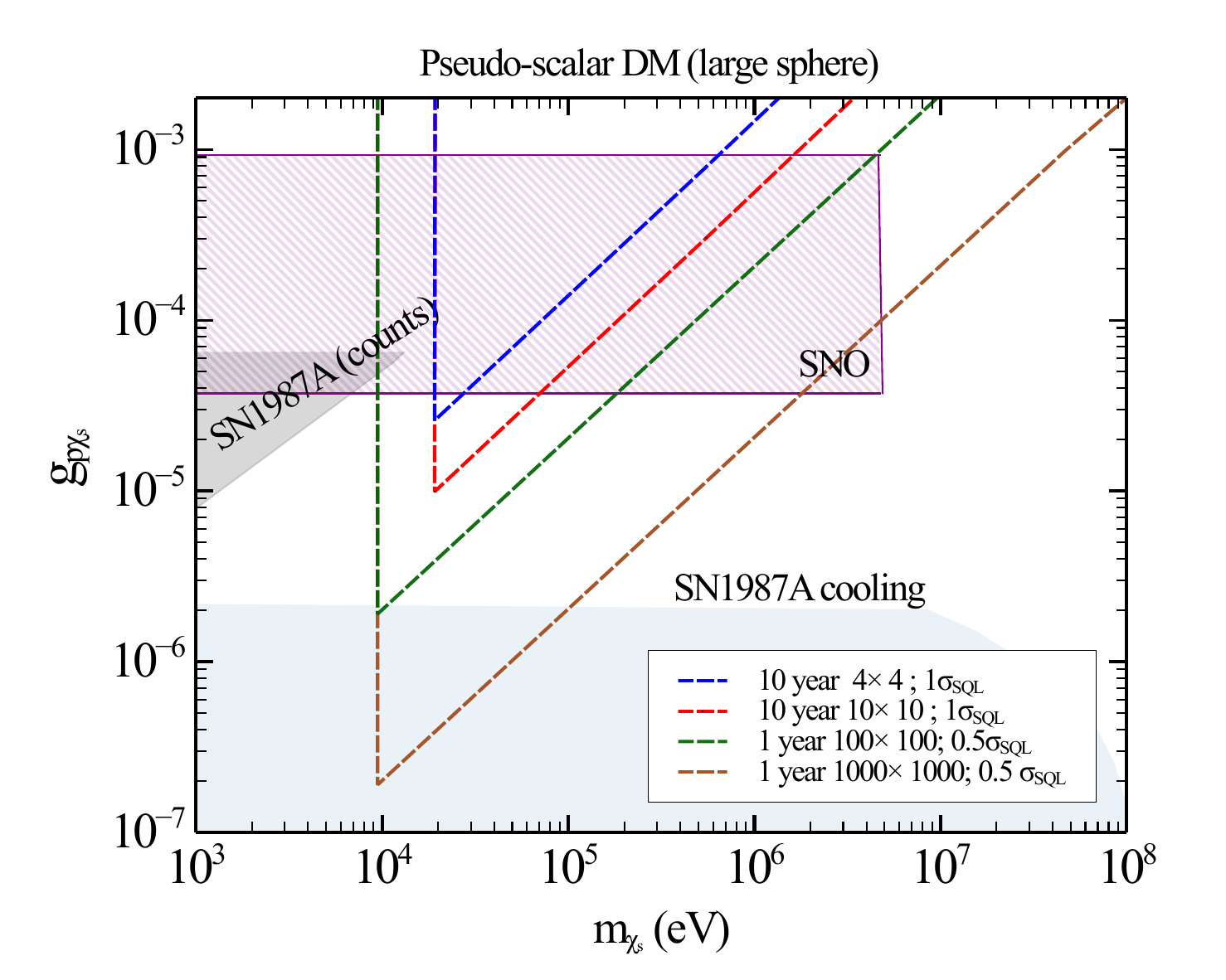}
    \caption{Projected sensitivity at $90\%$ C.L on the $m_{\chi_s}-g_{p\chi_s}$ plane using different combinations of large (diameter $ 200 $ nm) nanosphere array. 
    Shaded regions are existing constraints discussed in the text. }
    \label{fig:alp_dm_l}
\end{figure}

 We present the sensitivity at $90\%$ C.L. for the 200 nm diameter levitated sphere setup in Fig.\ref{fig:alp_dm_l}.
The limits with a $4\times 4$ array and a $10\times 10$ array with an integration time of $10$ years and detection threshold $q_{th}=1 \sigma_{\rm SQL}$ are shown by blue dashed and red dashed lines, respectively.
The limits with the $100\times 100$ and $1000\times 1000$ array with detection threshold $q_{th}=0.5 \sigma_{\rm SQL}$ are shown by green dotted lines and brown dotted lines, respectively.
For these two limits, we consider a year-long integration time.
The other relevant constraints such as SNO \cite{Bhusal:2020bvx}, SN1987A counts \cite{Lella:2023bfb} and SN1987A cooling \cite{Carenza:2019pxu} are also displayed for comparison.
The $4\times 4$ array with an integration time of $10$ years and detection threshold $q_{th}=1 \sigma_{\rm SQL}$ can probe $g_{p \chi_s}\sim 3\times 10^{-5}$ for $m_{\chi_s}\sim 10$ keV and can constrain the parameter space that is not yet excluded, as shown in Fig.\ref{fig:alp_dm_l}.
A $100\times 100$ array can constrain even up to $g_{p\chi_s}\sim 10^{-6}$ which is beyond the sensitivity of SNO \cite{Bhusal:2020bvx}.
Finally, an optimistic $1000\times 1000$ array search can even probe the region complementary to the bound from SN1987A cooling \cite{Carenza:2019pxu}. Since the framework is analogous to the ALP-nucleon interaction, the constraints applicable to ALPs (on $g_{ap}$) are also relevant here.
The sharp decrease in the sensitivity limits at low masses occurs when $q$ 
falls below the detection threshold. As previously shown, for ALP DM absorption, 
$q \sim m_a $, consequently, the detection threshold imposes a lower mass cutoff for this setup.
Similar constraints can also be obtained for sub-MeV scalar and vector DM. However, most of the parameter space in this range is ruled out by the existing experiment for these types of DM~\cite{An:2014twa}. 

{\bf II.  Small nanosphere ($ q \lesssim {\cal O}(10)~{\rm keV} $)}:
\label{sec:small}
Now we turn our attention to the {\it small} sphere with diameter $15$ nm ($r_{sp}=7.5$ nm). The differential recoil rate and the structure factor $S(q)$ can be obtained by our earlier discussion.
Such a small sphere with low mass trapped with frequency $\omega=2\pi \times 1$ kHz has SQL $\sigma_{\rm SQL}\sim 85$ eV for momentum impulses. Such a small threshold can help to detect particles with very low energy. 
It may seem that such a sphere with only $10^6$
nucleons (large sphere contains $10^9$) will reduce the target size. 
However, if the transferred momenta of the incoming particle $q\lesssim 2\pi/r_{sp}$ the coherence scattering takes place throughout the sphere, leading to a sensitivity enhanced by the factor $\sim N_p^2$. This can be seen in the second part of $S(q)$. For example, we present a few scenarios in the following subsections.

II.A {\it Sub-keV pseudo-scalar DM :}
We continue our earlier discussion about pseudo-scalar DM with nucleon coupling and consider the sub-keV mass regime.
\begin{figure}[!tbh]
    \centering
    \includegraphics[scale=0.35]{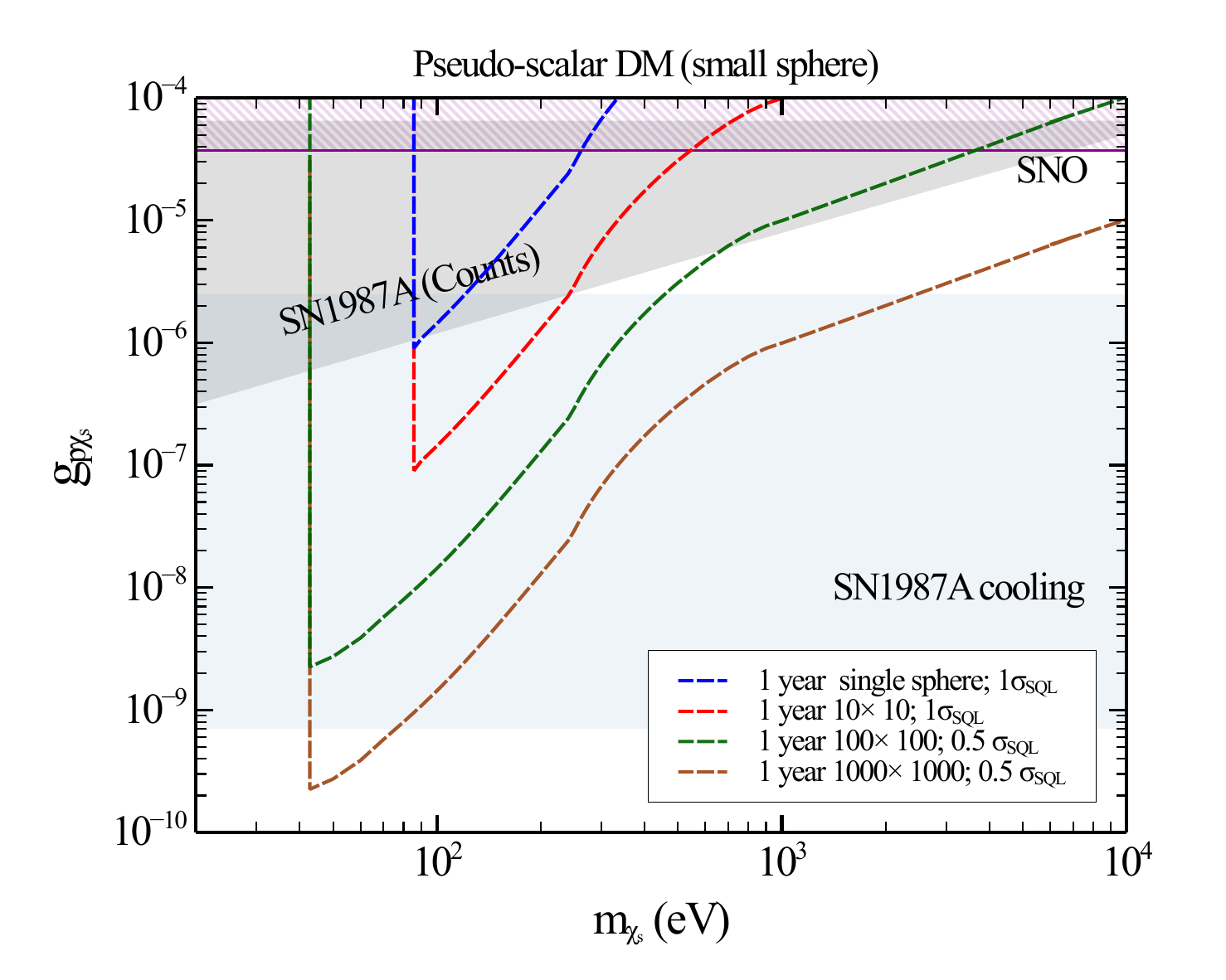}
    \caption{Projected sensitivity at $90\%$ C.L on the $m_{\chi_s}-g_{p\chi_s}$ plane using different combinations of  small (diameter $15$ nm) nanosphere array with $1$ year integration time.}
    \label{fig:alp_dm}
\end{figure}
In Fig.\ref{fig:alp_dm} we show the projected limits with a 15-nm diameter setup.
To denote the limits from various combinations of nanosphere arrays with 1 year integration time ( single sphere (blue dashed), $10\times 10$ (red dashed), $100\times 100$ (green dotted), $1000\times 1000$ (brown dotted)).
With a single 15-nm diameter levitated sphere set up, one can probe ALP DM up to $m_{\chi_s}\sim 85$ eV and $g_{p\chi_s}\sim 10^{-6}$ which is beyond the sensitivity of current direct detection experiments.
The setup can go beyond the exclusion regions of SNO \cite{Bhusal:2020bvx} and SN1987A counts \cite{Lella:2023bfb} using only a $10\times 10$ array with $q_{th}=1\sigma_{SQL}$.
With even larger array sizes such as $100\times 100$ and $1000\times 1000$, achieving $q_{th}=0.5 \sigma_{SQL}$ can lead to a substantial improvement in the sensitivity probing even up to $m_{\chi_s}\sim 50 $ eV.
Again, the threshold momentum for the small nanosphere is reflected in the low-mass cut-off. Also note that smaller spheres can probe even larger masses though the limits for $m_a \gtrsim 10^4$ eV will be weaker than those obtained with large spheres for the reason already explained before.

Some of the limits in Fig.\ref{fig:alp_dm} go through the parameter space already excluded by supernova constraints. However, as mentioned earlier, the supernova constraints are model dependent and are currently being questioned \cite{Bar:2019ifz}.
Hence, the region should be explored as a complementary probe in alternate terrestrial DM searches.
Thus DM searches with a levitated sphere can explore the previously  uncharted  region ($g_{p\chi_s}\gtrsim10^{-3}$), thus
opening new avenues for the phenomenology of BSM.

\begin{figure}[!tbh]
    \centering
    \includegraphics[scale=0.35]{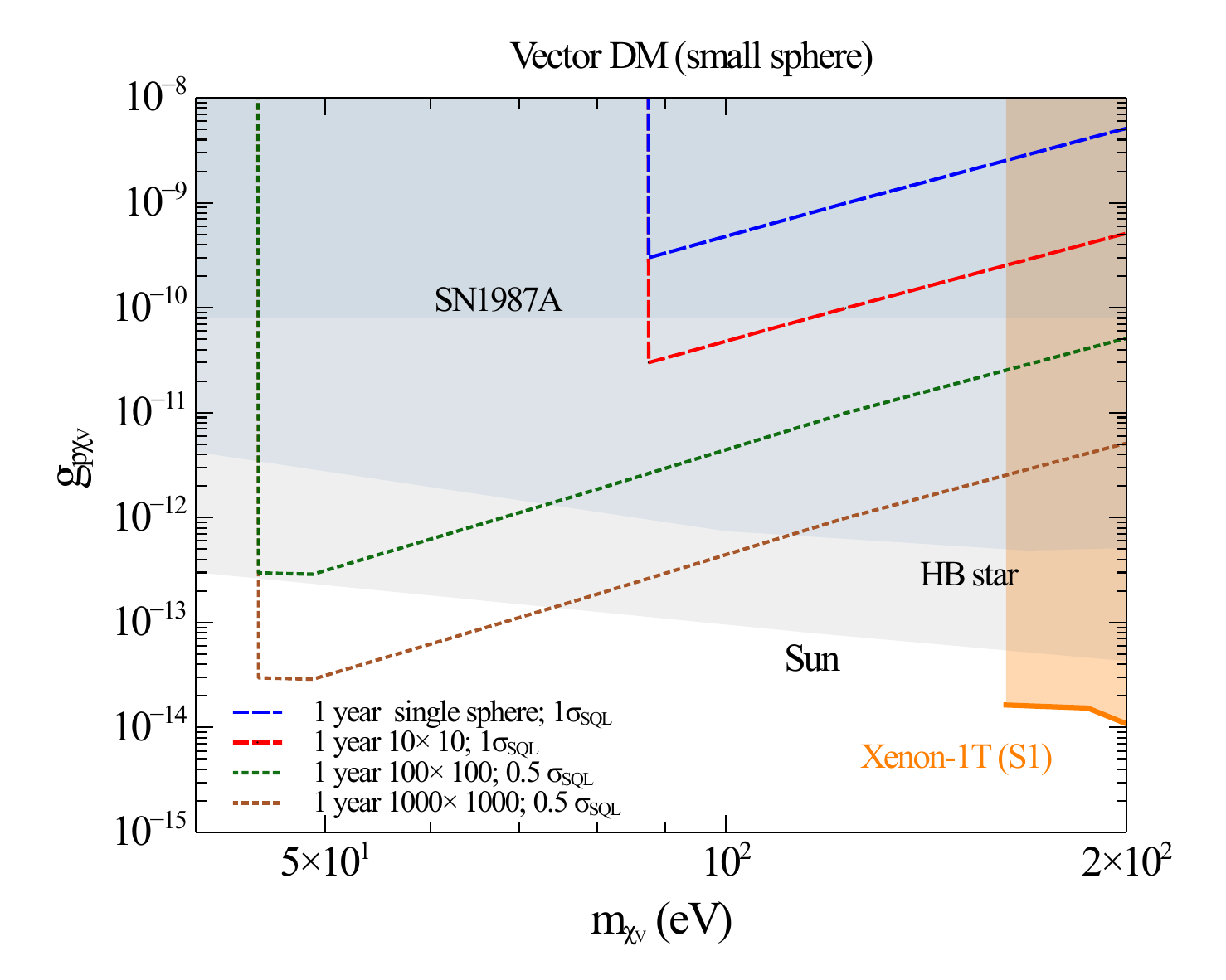}
    \caption{Projected sensitivity at $90\%$ C.L on the $m_{\chi_{_V}}-g_{p{\chi_{_V}}}$ plane using different combinations of 
    small (diameter $15$ nm) nanosphere array with $1$ year integration time.} 
    \label{fig:vec_dm}
\end{figure}

II.B {\it Sub-keV vector DM:}
\label{sec:vec_dm}
 Dark matter can also be a vector-like particle \cite{Hambye:2008bq} that has the
following effective interaction with the nucleon with mass $m_{\chi_V}$ and coupling strength $g_{p\chi_{V}}$: 
$\mathcal{L}_{V p}\supset -i g_{p{\chi_{_V}}}\left [\Bar{p}\gamma_\mu p\right ] \chi^\mu_{_V} $.
Following the previous discussion, we also show the bounds obtained for this DM through nucleon scattering in Fig.\ref{fig:vec_dm}.
The limits of various combinations of nanosphere arrays follow the same convention as used in Fig.\ref{fig:alp_dm}.
With a single 15nm diameter levitated sphere setup with $q_{th}=1 \sigma_{SQL}$ one can probe the vector DM up to $m_{\chi_{_V}}\sim 85$ eV and $g_{p{\chi_{_V}}}\sim 10^{-10}$. With larger array sizes such as $100\times 100$ and $1000\times 1000$ achieving $q_{th}=0.5 \sigma_{SQL}$ it can explore previously unconstrained regions. 
We also show the existing bound from loop-induced electron recoil from Xenon-1T \cite{XENON:2020rca} as well as astrophysical bounds from Sun \cite{Hardy:2016kme}, HB star \cite{Hardy:2016kme}, and SN1987A \cite{Chang:2016ntp}.
Apart from the astrophysical constraints, our analysis provides the first direct constraint on
vector dark matter-nucleon coupling in the sub-keV mass regime.

II.C {\it Earth-bound fermion DM :}
The ongoing multi-ton DD experiments apply strong constraints on the parameter space of weakly interacting DM. 
However, these experiments have a blind spot when it comes to strongly interacting DM, which remains undetectable by conventional search methods~\cite{Zaharijas:2004jv}.
As mentioned earlier DM particles with very high cross-section ($\sigma_{\chi N}\gtrsim 10^{-28}$ cm$^2$, $\chi$ is the fermion DM) with nucleons are expected to scatter with the earth's crust and atmospheres and eventually thermalize with the surrounding matter \cite{Leane:2022hkk,Bramante:2022pmn}.
Although only a tiny fraction $f_\chi$ of the total density of DM can have such large cross sections, their accumulation within the earth over a very large timescale can significantly enhance the density \cite{Bramante:2022pmn,McKeen:2023ztq}. 
 Despite their high number density, these DM particles ($\chi$) are associated with very small velocity ($\sim 10^{-6}/\sqrt{m_\chi/{\rm GeV}}c$) and thus can escape conventional DM searches in large detectors \cite{Zaharijas:2004jv}.
Earth-bound DM has the highest density on the surface around $m_\chi \sim$ GeV \cite{Bramante:2022pmn} and has momentum
$q\lesssim \mathcal{O}(10)$ keV. Hence, optically trapped {\it small} spheres become an ideal probe for such a scenario.

To calculate the recoil rate of such DM particles, the number density $n_\chi$ is obtained using a simplified approach. For $m_\chi\gtrsim 1.5$ GeV the evaporation rate is negligible \cite{Neufeld:2018slx}. Thus total DM no. bound on earth are given as $N_\chi \approx \Gamma_{\rm cap} t_{\oplus}$, where $\Gamma_{\rm cap}$ and $t_\oplus$ signify respectively the DM capture rate and age of earth \cite{Neufeld:2018slx}. 
For estimating the bounds, we have assumed the following approximations. 
Firstly, $\Gamma_{\rm cap}$ is given as $\sim f_c \Gamma_{\rm geo}$, where $\Gamma_{\rm geo}$ is the maximum geometric capture rate in the volume of the earth and $f_c$ is the capture fraction. For $M_\chi\gtrsim {\cal O}({\rm GeV})$ and $\sigma_{\chi n}\gtrsim 10^{-32}{\rm cm}^2$, $f_c$ can be taken as constant ($f_c=0.1$) to obtain a conservative limit \cite{Bramante:2022pmn}.  
\begin{figure}[!tbh]
    \centering
    \includegraphics[scale=0.34]{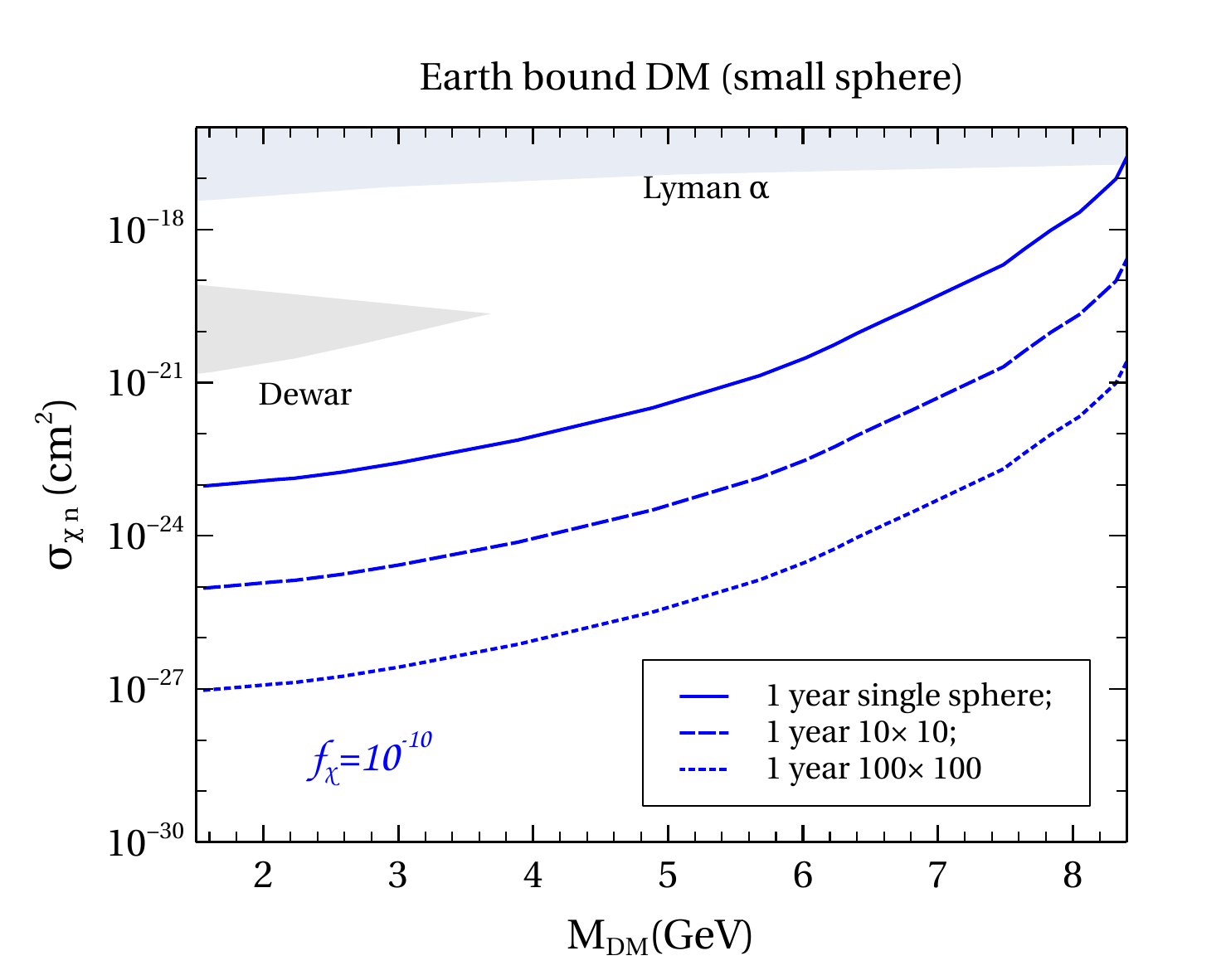}
    \caption{Projected sensitivity at $90\%$ C.L on the $m_{\rm DM}-\sigma_{\chi n}$ plane using different combinations of  
    small (diameter $15$ nm) nanosphere array with $1$ year integration time.}
    \label{fig:earth_small}
\end{figure}

In Fig.\ref{fig:earth_small} we show the constraints obtained only for the benchmark point: $f_\chi=10^{-10}$ (blue lines). These constraints are shown for different configurations of nanosphere arrays with varying line styles:  
(a) 1 single sphere with 1 year (solid) (b) $10\times10$ array with 1 year (dashed) (c)$100\times 100$ array with 1 year (dotted).
For this analysis, we have assumed that the threshold is $1\sigma_{SQL}$.
For the chosen benchmark, we have also portrayed the existing constraints of Lyman$\alpha$~\cite{Buen-Abad:2021mvc}, Dewar heating~\cite{Neufeld:2018slx,Neufeld:2019xes}.
Our bounds for  1-year single sphere and the $100\times 100$ array of 1 year are stronger than the existing constraints. 
Thus, an optically trapped sphere can provide 
direct detection of Earth-bound DM
and such limits can act as complementary to the bounds obtained through annihilation\cite{McKeen:2023ztq} or nuclear accelerators \cite{McKeen:2022poo}.

{\it Background.  }In our analysis we assumed the background to be negligible.
Compared to the large $\sim$ton size detectors, the recoil rates due to neutrinos are
highly suppressed due to their smaller cross-section.
For example, with a single large nanosphere the expected rate for neutrinos would be $\sim 10^{-19}$/year. Since neutrinos have energies $\gg 100$ eV, the coherence effect does not occur for smaller spheres, leading to a suppressed rate $\sim 10^{-22}$/year.
Low-energy secondary particles from 
high energy radiation can be distinguished by placing conventional particle detectors around the setup \cite{Afek:2021vjy}. 
The remaining background from the gas collision can be controlled for a large (small) nanosphere with smaller temperature $\simeq 4$K ($300$K)and vacuum pressure $\simeq 10^{-12}$mbar ($10^{-15}$mbar) 
at  SQL~\cite{Afek:2021vjy}. Finally, other noise-induced recoils can be easily
isolated from DM or ALP signals using directional sensitivity and correlations within large arrays~\cite{Monteiro:2020wcb}.

{\it Conclusion:}
In this letter, we have proposed an alternative approach to search the ALP nucleon coupling with the solar ALP of $14.4$ keV, different types of sub-GeV DM and Earth-bound DM.
Conventional detection methods face challenges in observing small nuclear recoils, resulting in reduced sensitivity to nuclear interactions. 
In contrast, optically trapped nanospheres offer the potential to probe previously unconstrained parameter space of solar-ALP and DM even
with a $4\times 4$ array of larger nanospheres ($200$ nm) which can be improved further with larger array sizes.
Using smaller spheres ($\sim$15 nm) significantly improves sensitivity, with a reduced target size compensated for by enhanced coherence. Larger arrays of such spheres allow for the probing of even smaller couplings, and the development of such arrays has already been demonstrated \cite{Manetsch:2024lwl}.
Arrays of nanospheres with varying sizes enable exploration of complementary mass and energy regimes. The large spheres are most sensitive to $q\gtrsim \mathcal{O}(10)$ keV, while the small spheres achieve maximum sensitivity at
$q\lesssim \mathcal{O}(100)$ eV. In the future, the incorporation of a third set of nanospheres with intermediate sizes, such as those with a diameter of 100 nm, could optimize the sensitivity in the
$q\sim \mathcal{O}(1)$ keV range. This development would improve the detection limits for solar ALP and Earth-bound DM. Thus, levitated optomechanical sensors offer a promising avenue to explore ALPs and different DM scenarios.

{\bf Acknowledgement.}
The authors thank David C. Moore and Gadi Afek for helpful discussions. 
SJ would like to thank Filippo Sala, Oleg lebedev, Genevieve Belanger, Sedric Delaunay, Jérémie Quevillon, Laura L. Honorez for the useful discussion and comments.
SJ also thanks Adrian Thompson for useful email conversations. The work of BD is supported by the U.S.~Department of Energy Grant DE-SC0010813.
The work of SJ is funded by CSIR, Government of India, under the NET JRF fellowship scheme with Award file No. 09/080(1172)/2020-EMR-I.

\bibliography{ref}

\newpage
\appendix
\section{ALP nucleon scattering}
\begin{figure}[!tbh]
    \centering
    \includegraphics[scale=0.45]{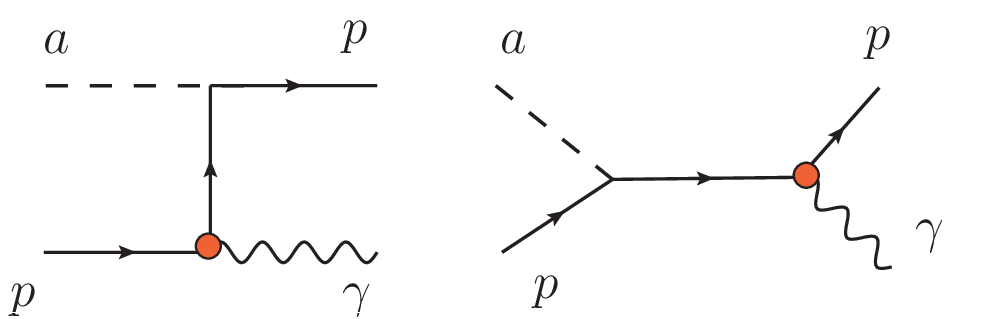}
    \caption{Feyman diagram of ALP proton scattering}
    \label{fig:diag_aN}
\end{figure}
ALP can interact with protons (neutrons) by the process as described in Fig.\ref{fig:diag_aN} and 
the amplitude of the process $a (k_1)+N(k_2)\to \gamma (k_3)+N (k_4)$ is given by
\cite{Guo:2022hud},
\bew
\begin{eqnarray}
    \mathcal{M}&=&i g_{ap} \left[ -\Bar{u}(k_3) \gamma_5 \frac{1}{\not{k_3}-\not{k_4}-m_N} \Gamma_N~ u(k_2)-\Bar{u(k_3)}~\Gamma_N \frac{1}{\not{k_1}+\not{k_2}-m_N} \gamma_5 u(k_2) \right]
    \label{eq:amp}
\end{eqnarray}
where,$\Gamma_N= \left[e \gamma^\mu + \frac{\kappa}{4 m_N} \sigma^{\mu \nu}{k_{4}}_\nu\right] \epsilon_\mu(k_4)$, $m_p$ and $q$ are the mass of the proton and the recoil momentum of the target. The averaged amplitude square is given by,
\begin{eqnarray}
     |\mathcal{M}|^2&=& \left[ M_1^2+M_2^2+M_3^2 \right],~{\rm where,}\\
     M_1^2 &=& \frac{4 m_p^2(3 m_a^2 +E_a m_p)-4(m_a^2+2 E_a m_p)q^2+\frac{\kappa^2}{m_p^2} (m_p^2 - q^2)^2 (2 E_a m_p - m_p^2 + q^2)}{{(q^2-m_p^2)^2}}\\
     M_2^2 &=& \frac{ m_p^2(3 m_a^2+2 E_a m_p)-(m_a^2+2 E_a m_p)q^2}{E_a^2 m_p^2} \nonumber\\
     &&-\frac{\frac{\kappa^2}{
  m_p^2} (m_a^2 + 2 E_a m_p) (-4 E_a^2 m_p^2 + m_a^2 (m_p^2 - q^2) + 
   2 E_a m_p (m_a^2 + m_p^2 - q^2))}{E_a^2 m_p^2}\\
    M_3^2 &=& -8\frac{m_a^2+m_p^2-q^2}{m_p^2-q^2} +\frac{ \frac{\kappa^2}{m_p^2}   (m_a^2 - 2 E_a m_p) (m_a^2 + 2 E_a m_p - m_p^2 + q^2)}{(m_p^2-q^2)}\\
    \end{eqnarray}
    In the limit $q\ll m_p$ this can be further simplified as,
   \begin{eqnarray*}
    |\mathcal{M}|^2&\approx&\frac{e^2 g_{ap}^2}{2} \left[ \frac{4 m_p^2(3 m_a^2 +E_a m_p)-4(m_a^2+2 E_a m_p)q^2}{(q^2-m_p^2)^2}+\frac{ m_p^2(3 m_a^2+2 E_a m_p)-(m_a^2+2 E_a m_p)q^2}{E_a^2 m_p^2} -8\frac{m_a^2+m_p^2-q^2}{m_p^2-q^2}\right]
    \end{eqnarray*}
\eew
The differential cross-section can be obtained by,
\begin{eqnarray}
    \frac{d\sigma}{dq}=\frac{2q}{64 \pi s}\frac{1}{(p^i_{cm})^2} |\mathcal{M}|^2
\end{eqnarray}
where, $s=(k_1+k_2)^2$ is the Mandelstam variable and $p^i_{cm}$ is given by, 
\begin{eqnarray*}
   (p^i_{cm})^2&=&\frac{1}{4s}(s-(m_a+m_p)^2)(s-(m_a-m_p)^2) \\
\end{eqnarray*}
This can be further simplified in the limit $m_a\ll q$ as, 
\begin{eqnarray}
    \frac{d\sigma}{dq}=\frac{2q}{16 \pi (2 m_p E_a)^2}\frac{e^2 g_{a p}^2}{2}\left[ \frac{8 m_p E_a}{q^2-m_p^2}+\frac{ m_p^2-q^2}{E_a m_p} -8\right].
\end{eqnarray}

\section{Earth bound DM recoil}
The total captured DM no.  $N_\chi$ in earth evolves as,
\begin{equation}
    \frac{dN_\chi}{dt}=\Gamma_{\rm cap}- N_\chi \Gamma_{\rm evap},
\end{equation}
where, $\Gamma_{\rm cap},\Gamma_{\rm evap}$ signify capture and evaporation rate respectively. For this analysis, we ignore the DM annihilation.
Thus the total captured DM no. throughout earth's age is given by,
\begin{equation}
    N_\chi = \frac{\Gamma_{\rm cap}}{\Gamma_{\rm evap}} \big(1-e^{-\Gamma_{\rm evap} t_{E}} \big),
\end{equation}
where, $t_E=10^{17}$ sec is the age of earth.
$\Gamma_{\rm evap}$ is negligible for $M_\chi\gtrsim 1.5$ GeV and $\sigma_{\chi N}\gtrsim 10^{-27}$cm$^2$ \cite{Neufeld:2018slx}. 
Capture rate is given by $\Gamma_{\rm cap} =f_c \Gamma_{\rm geo}$, where $\Gamma_{\rm geo}$ is the rate of DM particle encountered by earth approximated as geometric cross section of earth multiplied by local DM flux \cite{McKeen:2022poo}, 
\begin{equation}
    \Gamma_{\rm geo} = \pi R_\oplus^2 \frac{\rho_\chi}{m_\chi} v_{\rm eff},
\end{equation}
$R_\oplus, v_{\rm eff}$ are  radius of earth and local effective DM velocity ($\approx 250$ km/s) respectively \cite{McKeen:2022poo}.

\begin{figure}[]
    \centering
    \includegraphics[scale=0.35]{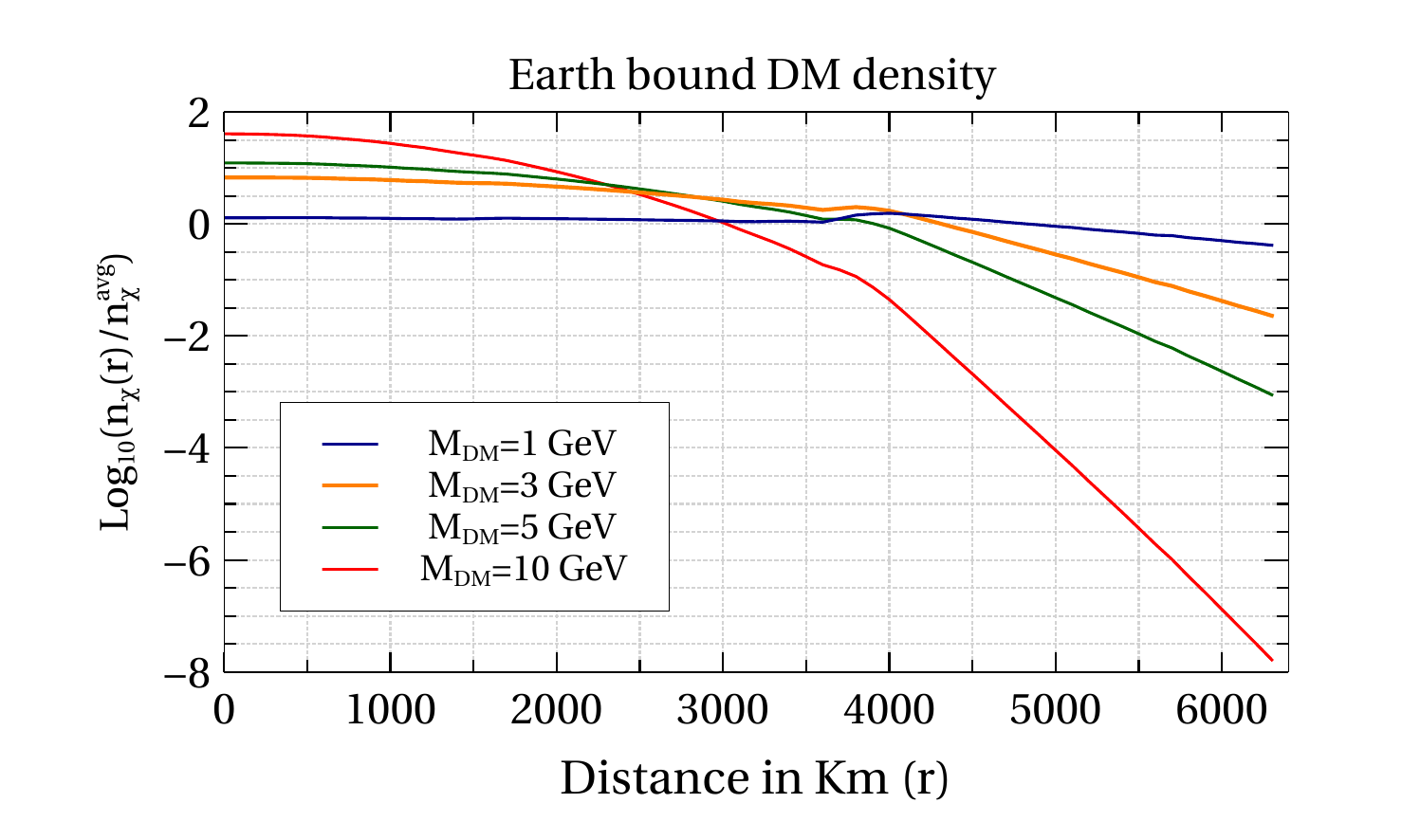}
    \caption{Variation of number density of captured DM particles relative to the average number density with the distance from center for different DM masses $M_\chi=1$ GeV (blue), $3$ GeV (orange), $5$ GeV (green) and $10$ GeV (red) depicted by different colors. }
    \label{fig:den}
\end{figure}

Then the average DM number density in earth is given by, $n_\chi^{\rm avg}=3 N_\chi/(4 \pi R_\oplus^3)$.
However, due to the Earth's gravitational potential the DM no. density is not uniform, rather for heavier mass DM particles tend to accumulate towards the earth's core.
This can be evaluated as analogous to dilute gas particles in a spherically symmetric potential $V(r)$.
The DM pressure density $p(r)=n_\chi (r)  T$ (in natural units) is given by
    $p(r)\propto \exp{(-V(r)/T)}$
where $V(r)$ is the gravitational potential of each DM particle at radius $r$.
Assuming a thermal equilibrium, 
we numerically evaluate $p(r)/p(0)$, $p(0)$ being the density at radius $r=0$, using the density profile in ref.\cite{Neufeld:2018slx} and the temperature profile in ref.\cite{Dziewonski:1981xy}. From the pressure density, one can easily evaluate the DM density $n_\chi(r)$ in terms of $n(0)$, using which the ratio $n_\chi(r)/n_\chi^{\rm avg}$
can be obtained.
The numerical result is shown in Fig. \ref{fig:den}.

In Fig.\ref{fig:den} we show the variation of the captured DM density in terms of $\log_{10}(n_\chi(r)/n_\chi^{\rm avg})$ with the distance $r$ (in Km) from the center of the Earth.
The results are shown for four different DM masses $M_\chi=1$ GeV (blue), $3$ GeV (orange), $5$ GeV (green) and $10$ GeV (red) depicted by different colors.
Note that lighter DM particles $M_\chi\lesssim 1$ GeV are distributed almost uniformly at all distances from the center, whereas heavier DM particles tend to accumulate toward the center of the Earth.
As a result, one can observe a significant decrease in the exact number density $n_\chi (R_\oplus)$ on the surface of the earth compared to the average number density.

\begin{figure}[!tbh]
    \centering
    \includegraphics[scale=0.35]{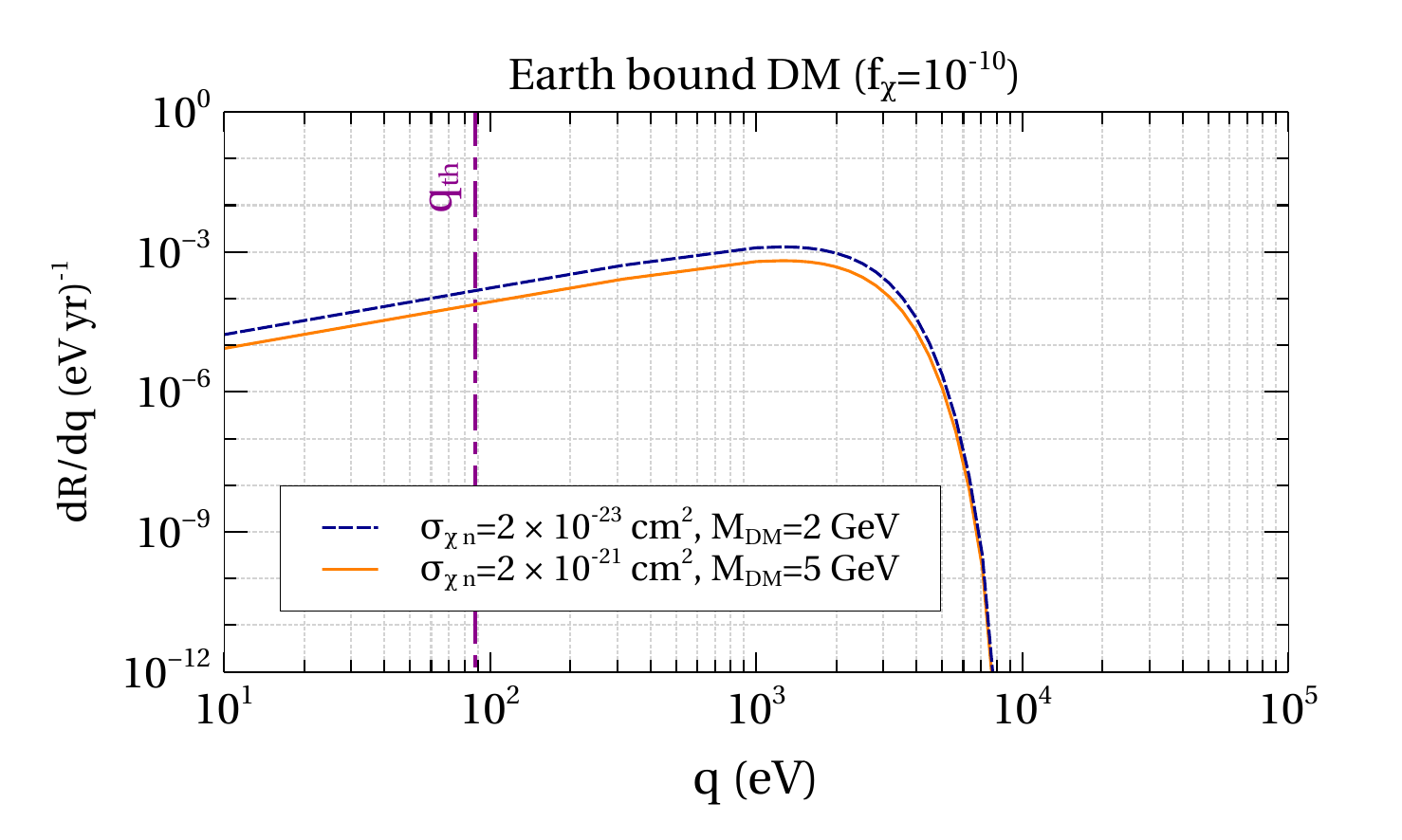}
    \caption{Expected recoil rate in a single 15 nm diameter nanosphere with $f_\chi=10^{-10}, f_c=0.1$ for two benchmark points $M_\chi=2$ GeV $\sigma_{\chi n}=2 \times 10^{-23}$ cm$^2$;
and $M_\chi=5$ GeV $\sigma_{\chi n}=2 \times 10^{-21}$ cm$^2$. }
    \label{fig:erth_rate}
\end{figure}

The earth bound DM velocity distribution is given by,
\begin{equation}
    f_\chi^{\rm bound}=\frac{1}{\mathcal{N}}\exp{\bigg(-\frac{v^2}{v_{th}^2} \bigg)} \Theta(v_{\rm esc}-v_{\rm th}),
\end{equation}
where, $v_{th}=\sqrt{\frac{8T_\chi}{m_\chi}}$ is the local thermal velocity, $T_\chi\sim 0.01$eV (300 K) is the earth's temperature at  surface.
$v_{\rm esc}$ is the earth's escape velocity $\sim 11$ km/s in SI units.
$\mathcal{N}$ is the overall normalization constant.
The expected recoil rate in a single 15 nm diameter nanosphere is shown in Fig.\ref{fig:erth_rate} with $f_\chi=10^{-10}, f_c=0.1$ for two benchmark points $M_\chi=2$ GeV $\sigma_{\chi n}=2 \times 10^{-23}$ cm$^2$;
and $M_\chi=5$ GeV $\sigma_{\chi n}=2 \times 10^{-21}$ cm$^2$. 

\end{document}